\documentclass{jaa}
\usepackage{graphicx}
\usepackage{natbib}
\usepackage{amsmath}
\usepackage{subcaption}
\usepackage{xcolor}
\usepackage{booktabs}
\usepackage{url}
\usepackage{newtxtext,newtxmath}
\usepackage{stfloats}
\usepackage{placeins}
\usepackage{hyperref}
\FloatBarrier

\bibliography{references} % Points to references.bib

\begin{document}\sloppy

%%paper title
%%For line breaks \\ can be used within title
\title{\textbf{Multi-Band Optical Variability Characteristics of the $\gamma$-ray emitting blazar 1ES 0647+250 during its 2020 flare}}

%%author names are separated by comma (,)
%%use \and before the last author name
%%use a * along with the number separated by comma
%% for the  author for correspondence
%%\textsuperscript{number} is used for affiliation
%%\affilOne, \affilTwo etc., upto \affilTwentyfive is possible
%%Please note the first letter after \affil is capitalised in the command
%%

\author{K.Subbu Ulaganatha Pandian\textsuperscript{1,6*}, 
C. S. Stalin\textsuperscript{2},
S. Muneer\textsuperscript{2},
S. Umayal\textsuperscript{2},
Aditi Agarwal\textsuperscript{3},
Amit Kumar Mandal\textsuperscript{4}, 
and B. Natarajan\textsuperscript{5}}
\affilOne{\textsuperscript{1}Research and Development Centre, Bharathiar University, Coimbatore - 641046, India\\}
\affilTwo{\textsuperscript{2}Indian Institute of Astrophysics, Block II, Koramangala, Bangalore-560034, India\\}
\affilThree{\textsuperscript{3}Centre for Cosmology and Science Popularization, SGT University, Gurugram 122505, India \\}
\affilFour{\textsuperscript{4}Center for Theoretical Physics, Polish Academy of Sciences, Al. Lotnik\'ow 32/46, 02-668 Warsaw, Poland\\}
\affilFive{\textsuperscript{5}Government Arts and Science College, Sivakasi – 626124 , India \\}
\affilSix{\textsuperscript{6}Government High School, Sirumalaipudur – 624003 , India \\}

%%escape two column mode for title, affiliation and abstract
%%by giving \twocolumn command as shown

\twocolumn[{

\maketitle

%%include \corres to print the corresponding author Email id
\corres{subbuathoor@gmail.com}

%%include \msinfo for
%%manuscript information such as
%%received, revised and accepted dates
%%
\msinfo{23 July 2024}{ 18 September 2026}

%%abstract
\begin{abstract}
We present a multi-band optical variability study of the $\gamma$-ray emitting high-synchrotron-peaked blazar 1ES~0647+250 during its enhanced optical activity in 2020. The analysis utilizes 107 epochs of high-precision $B$, $V$, and $R$-band photometry obtained with the 75 cm telescope at the Vainu Bappu Observatory, combined with high-cadence $z_g$ and $z_r$ survey data from the Zwicky Transient Facility. The combined light curves reveal coherent variability across all optical filters, with the source reaching a peak $R$-band magnitude of 15.74 mag. We calculated the variability amplitude, finding the highest intrinsic variability in the B- band. Spectral analysis using color--magnitude diagrams reveals a statistically significant ``Bluer-When-Brighter'' trend, supported by a Pearson correlation coefficient of $r = 0.68$ ($p < 0.001$) for the (V$-$R) index. Furthermore, the discrete correlation function analysis indicates a near-zero time lag between the $V$ and $R$ bands, suggesting simultaneous emission within the observational cadence. The observed spectral hardening is consistent with synchrotron-dominated optical emission, likely associated with enhanced particle acceleration within the jet during the active phase of 1ES~0647+250.
\end{abstract}

%%insert keywords separated by 3 hyphens using \keywords{words}
\keywords{galaxies:active --- galaxies:jets --- BL Lacertae objects: individual: 1ES 0647+250}

}]
%%close the twocolumn escape here

%%include \doinum{number}for the DOI number in the header
%%include \volnum{number} for the volume number in the header
%%include \year{yyyy} for  year of publication in the header
%%include \pgrange{num--num} page range of article in the header
%%include \artcitid{num} for the article citation id
%%include \lp to print last page of the article
%%include \setcounter{page}{pagenum} for the exact starting page of the article

\doinum{12.3456/s78910-011-012-3}
\artcitid{\#\#\#\#}
\volnum{000}
\year{0000}
\pgrange{1--}
\setcounter{page}{1}
\lp{1}

\section{Introduction}

Blazars are a class of radio-loud active galactic nuclei (AGN) characterized by relativistic jets oriented close to the observer’s line of sight. They are broadly classified into two subclasses: BL Lacertae objects (BL Lacs) and flat-spectrum radio quasars (FSRQs), based on the absence or presence of strong emission lines in their optical spectra. While FSRQs exhibit prominent broad emission lines, BL Lac objects show nearly featureless optical spectra, with emission lines, if present, having equivalent widths smaller than 5~\AA. They exhibit rapid and extensive flux variability throughout the electromagnetic spectrum, spanning from low-energy radio to high-energy TeV $\gamma$-rays \citep{1995PASP..107..803U,1995ARA&A..33..163W}. In addition to flux variations, they also show polarization variations \citep{1980ARA&A..18..321A,2017ApJ...835..275R,2022MNRAS.510.1809P,2022MNRAS.517.3236R}. These observed characteristics of blazars are attributed to their relativistic jets being aligned closely with the observer's line of sight, with the observed radiation dominated by Doppler-boosted non-thermal jet emission. Optical variability studies of blazars have provided valuable insights into jet emission processes and variability characteristics across different timescales (e.g., \citealt{1995ARA&A..33..163W,2017ApJ...844...32P,2022JApA...43...48P}).

The spectral energy distribution of blazars is characterized by a double-peaked structure. The low-energy component, extending from radio to X-ray frequencies, is attributed to synchrotron emission from relativistic electrons in the jet. The high-energy component, spanning from X-ray to $\gamma$-ray energies, is commonly explained by inverse Compton scattering in leptonic models, although hadronic or lepto-hadronic scenarios have also been proposed \citep{2017SSRv..207....5R}.

Variability in blazars is observed over a range of timescales. Intraday variation or micro-variation involves flux changes of up to several tenths of a magnitude occurring within minutes to a day \citep{1995ARA&A..33..163W}. Short-term variation refers to fluctuations exceeding one magnitude over several months, while long-term variation encompasses flux changes spanning several magnitudes over months to years \citep{2001A&A...374..435M,2008AJ....135.1384G}. Various mechanisms have been proposed to explain flux variations in blazars, including accretion disk instabilities, jet precession, shock propagation within the jet, and variations in the Doppler factor caused by relativistic plasma motion \citep{1996ASPC..110...42W,2017Natur.552..374R}. In addition, flux variations are often accompanied by changes in color or spectra, providing insights into the source structure. The analysis of flux variability and spectral behavior provides important constraints on electron cooling dynamics within blazar jets and helps test theoretical models \citep{2008AJ....136.2359G,2003A&A...402..151R}.

In addition to flux variations, blazars often exhibit spectral variability, which can be investigated through color--magnitude analysis. The commonly observed Bluer-When-Brighter (BWB) trend suggests that the spectrum becomes harder when the source brightens, providing important clues about the evolution of the electron energy distribution and possible shifts in the synchrotron emission peak.

In this work, we investigate the multi-band optical variability properties of the $\gamma$-ray emitting high synchrotron peaked (HSP) blazar 1ES~0647+250 using observations obtained in the $B$, $V$, and $R$ bands with the 75~cm telescope at the Vainu Bappu Observatory (VBO), Kavalur, India, along with archival data from the Zwicky Transient Facility (ZTF).  The structure of this paper is as follows: Section~2 describes the source and observations, Section~3 outlines the data reduction procedures, Section~4 presents the results, Section~5 discusses the physical implications, and Section~6 summarizes the conclusions.

\section{Source and Observations}
\subsection{Source}

1ES~0647+250 is a HSP (\citealt{2002A&A.384.56C}) TeV $\gamma$-ray emitting blazar \citep{2011ApJ...742...43A,2022ApJS..263...24A} with an uncertain redshift. It has been detected across multiple wavelengths, including radio, optical, X-ray and $\gamma$-ray bands. Its broadband emission is dominated by non-thermal radiation from relativistic jet oriented close to the line of sight. From deep imaging of the host galaxy, \citet{2011A&A...534L...2K} estimated a redshift of $z = 0.41 \pm 0.06$. The source is known to be variable across the electromagnetic spectrum, including radio \citep{2014ApJ...797...25P}, optical \citep{2009MNRAS.398..832K,2018A&A.620A.185N}, X-rays \citep{2005ApJ...625..727P}, and $\gamma$-rays \citep{2016A&A...593A..98L,2023A&A...670A..49M}. 

\begin{figure}[!t]
\centering
\includegraphics[width=\columnwidth]{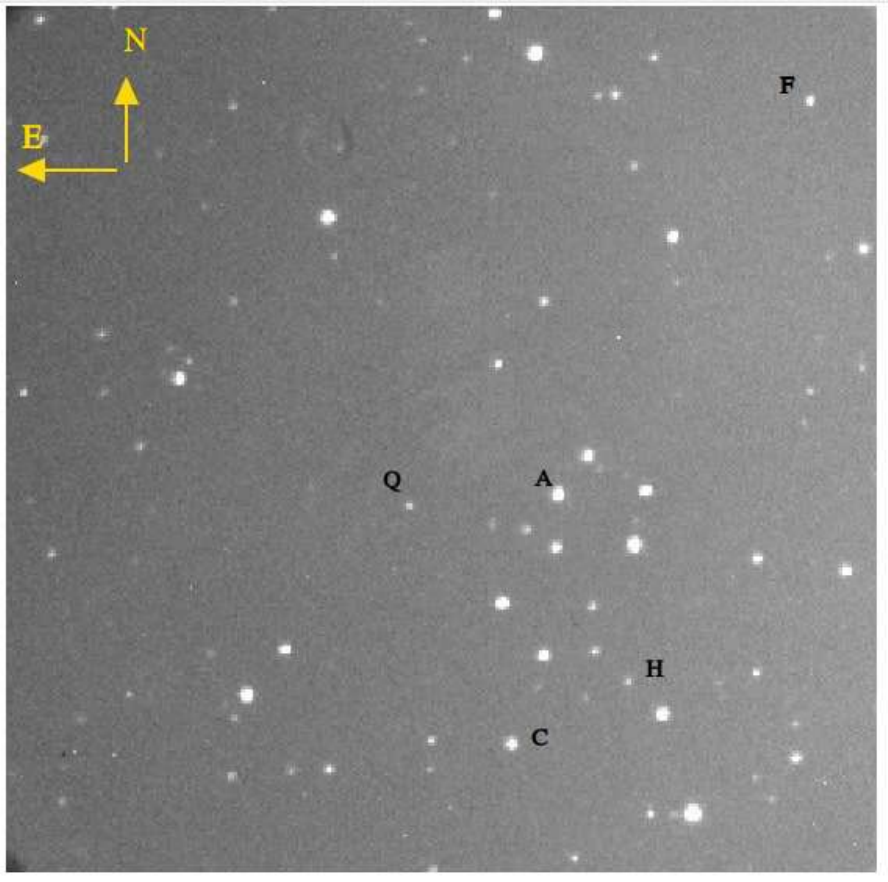}
    \caption{The observed field of 1ES~0647+250. Comparison stars used for differential photometry are identified and labeled alongside the target blazar.}
\label{figure-1}
\end{figure}

\begin{table}[h]
\centering
\caption{Log of the photometric monitoring observations}
\label{table-1}
\begin{tabular}{lccc} \hline
Date  &  \multicolumn{3}{c}{Number of frames} \\
           &  B & V  &   R \\ \hline
26-01-2020 & 0  & 1  &  2  \\
27-01-2020 & 0  & 1  &  1  \\
28-01-2020 & 1  & 1  &  2  \\
29-01-2020 & 1  & 2  &  3  \\
30-01-2020 & 2  & 3  &  2  \\
31-01-2020 & 0  & 1  &  2  \\
01-02-2020 & 0  & 2  &  2  \\
02-02-2020 & 0  & 2  &  1 \\
15-02-2020 & 2  & 2  &  2 \\
16-02-2020 & 2  & 2  &  3 \\
17-02-2020 & 2  & 1  &  1 \\
18-02-2020 & 2  & 2  &  3 \\
19-02-2020 & 3  & 3  &  2 \\
20-02-2020 & 1  & 3  &  2  \\
21-02-2020 & 1  & 2  &  2 \\
22-02-2020 & 1  & 1  &  1 \\
25-02-2020 & 1  & 2  &  0 \\
26-02-2020 & 0  & 2  &  2 \\
27-02-2020 & 0  & 1  &  1 \\
28-02-2020 & 0  & 1  &  2  \\
16-03-2020 & 1  & 1  &  2 \\
17-03-2020 & 1  & 2  &  2 \\
18-03-2020 & 2  & 1  &  1 \\
19-03-2020 & 1  & 2  &  1 \\ \hline 
\end{tabular}
\end{table}

In early 2020, 1ES~0647+250 exhibited enhanced activity across multiple wavelengths, providing an opportunity to investigate its jet properties through coordinated optical monitoring. Despite its strong multi-wavelength variability on longer time scales, several studies have reported an absence of significant intranight optical variability for this source \citep{2023MNRAS.524L..66N}. 

From long-term ($\sim$6 years) monitoring in the optical ($R$ band) and radio (15 GHz) regimes, \citet{2016A&A...593A..98L} and \citet{2018A&A.620A.185N} found a gradually increasing trend in both the light curves, with the optical data exhibiting additional flares superimposed on this trend. Cross-correlation analysis further revealed that the optical and radio variations are closely correlated, with the optical leading over the radio emission. Similarly, multi-wavelength variability studies by \citet{2023A&A...670A..49M} found that while optical and $\gamma$-ray flux variations are correlated with essentially zero lag, the radio variations lag behind both. 

Although 1ES~0647+250 has been characterized for long-term (yearly) and short-term (intranight) variability, there is currently a lack of information regarding its optical variability characteristics in multiple bands on day-like (intermediate) time scales. This study aims to address that gap using coordinated multi-filter observations during its 2020 active state.

\begin{figure*}[!t]
\centering
\includegraphics[width=\textwidth]{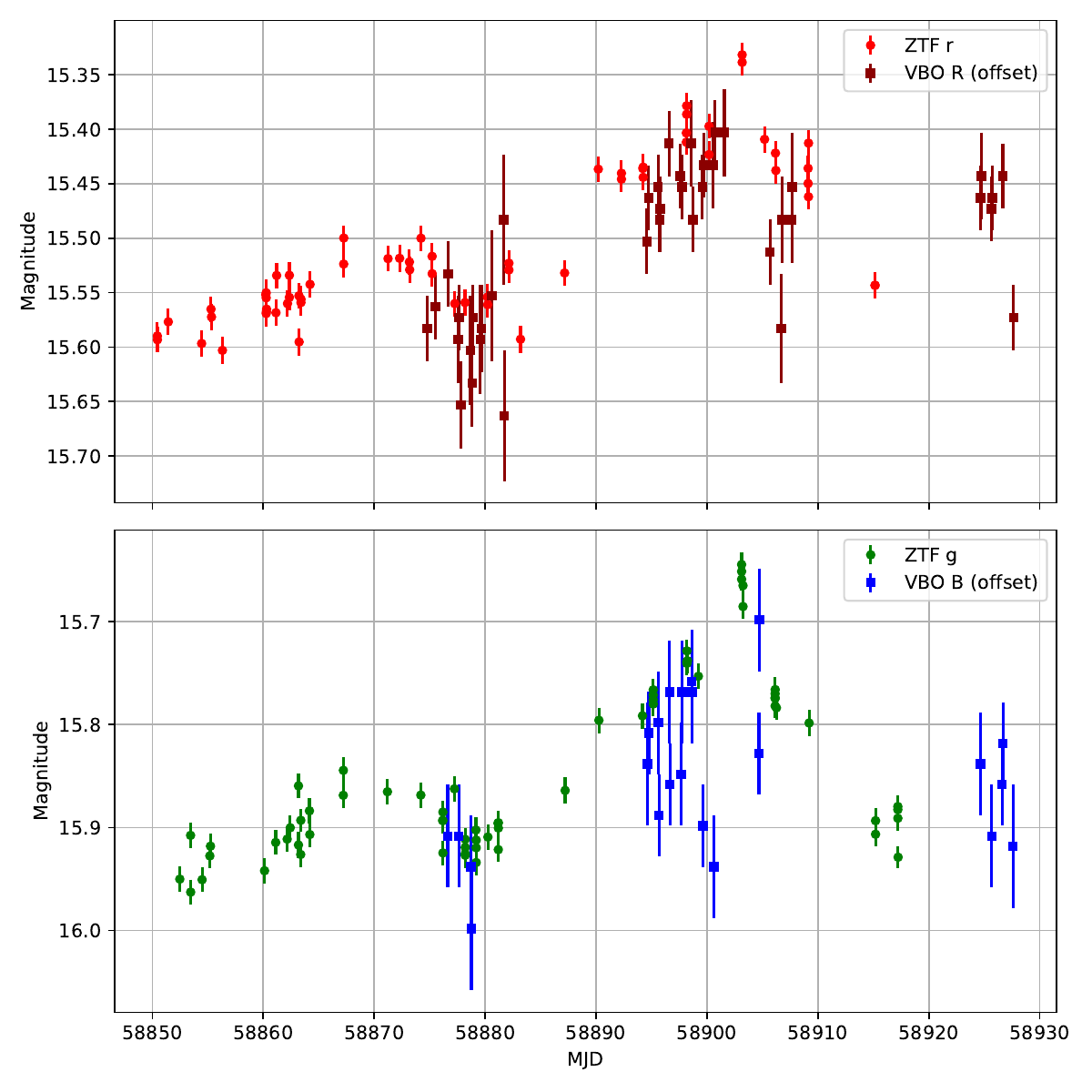}
    \caption{Optical light curves of 1ES 0647+250 from ZTF and VBO observations. The top panel shows ZTF $r$ and VBO $R$-band data, while the bottom panel shows ZTF $g$ and VBO $B$-band data. Constant offsets were applied to the VBO data to account for systematic differences between the instruments. Error bars denote $1\sigma$ uncertainties.}
\label{figure-2}
\end{figure*}
	
\subsection{Observations}
The optical monitoring of 1ES~0647+250 during its 2020 flaring activity was carried out using a combination of targeted observations and archival survey data, thereby enabling improved temporal coverage. This approach allowed us to capture both the rapid intra-night variations and the longer-term evolution of the source with enhanced continuity.

\subsubsection{\textbf{VBO Observations}}
Optical observations were carried out in the $B$, $V$, and $R$ filters using the 75 cm telescope at VBO, Kavalur, operated by the Indian Institute of Astrophysics. The telescope was equipped with a liquid nitrogen-cooled TEK CCD camera mounted at the $f/13$ Cassegrain focus. The CCD detector consists of $1024 \times 1024$ pixels, each with a pixel size of 24~$\mu$m, providing a scale of 6.8 arcsec mm$^{-1}$ (0.48 arcsec pixel$^{-1}$) and a field of view (FoV) of approximately $8 \times 8$ arcmin$^{2}$.

The seeing during the observations, derived from the median full width at half maximum (FWHM) of the stars in the frames, ranged between 3.0$''$ and 3.6$''$. The CCD was maintained at $-80^{\circ}$C to minimize dark current. The target field was selected to include multiple comparison stars for reliable differential photometry. 

Data were acquired over a total of 24 nights between January 26, 2020, and March 19, 2020. On each night, multiple frames were obtained in the $B$, $V$, and $R$ filters, with typical integration times of 2700~s, 2400~s, and 1200~s, respectively. The log of these observations is given in Table~\ref{table-1}.

\subsubsection{\textbf{ZTF Archival Data}}
To complement the VBO observations, we retrieved publicly available data from ZTF. We used the $z_g$ (green) and $z_r$ (red) filters, which provide high-cadence monitoring of the northern sky. Only data points with \texttt{catflags} equal to 0, indicating reliable detections, were selected for analysis. The ZTF magnitudes are reported in the AB photometric system, and the uncertainties correspond to the $1\sigma$ photometric errors produced by the ZTF data reduction pipeline \citep{2019PASP.131.068003B}.

\begin{table}[!t]
\centering
\caption{R-band photometry of 1ES 0647+250 from VBO. Here, $R$ and $\sigma_R$ are in magnitude units. }
\label{table-2}
\small
\begin{tabular}{ccc}
\toprule
MJD & $R$ & $\sigma_R$ \\
\midrule
58874.7736 & 15.79 & 0.03 \\
58875.5257 & 15.77 & 0.03 \\
58876.6639 & 15.74 & 0.03 \\
58877.5799 & 15.80 & 0.04 \\
58877.6618 & 15.78 & 0.03 \\
58877.8153 & 15.86 & 0.04 \\
58878.6799 & 15.81 & 0.05 \\
58878.8285 & 15.84 & 0.04 \\
58878.9146 & 15.78 & 0.03 \\
58879.5729 & 15.80 & 0.05 \\
58879.6736 & 15.79 & 0.04 \\
58880.6007 & 15.76 & 0.06 \\
58881.6639 & 15.69 & 0.06 \\
58881.7349 & 15.87 & 0.06 \\
58894.5764 & 15.71 & 0.03 \\
58894.7319 & 15.67 & 0.03 \\
58895.5813 & 15.66 & 0.03 \\
58895.7375 & 15.69 & 0.03 \\
58895.7854 & 15.68 & 0.03 \\
58896.5729 & 15.62 & 0.03 \\
58897.5868 & 15.65 & 0.03 \\
58897.7382 & 15.66 & 0.03 \\
58898.5715 & 15.62 & 0.04 \\
58898.7222 & 15.69 & 0.03 \\
58899.5743 & 15.66 & 0.03 \\
58899.7250 & 15.64 & 0.03 \\
58900.5701 & 15.64 & 0.04 \\
58900.7201 & 15.61 & 0.03 \\
58901.5688 & 15.61 & 0.04 \\
58905.6824 & 15.72 & 0.03 \\
58906.6754 & 15.79 & 0.05 \\
58906.7675 & 15.69 & 0.04 \\
58907.6539 & 15.69 & 0.04 \\
58907.6755 & 15.66 & 0.05 \\
58924.6410 & 15.67 & 0.03 \\
58924.7278 & 15.65 & 0.04 \\
58925.6417 & 15.68 & 0.03 \\
58925.7236 & 15.67 & 0.03 \\
58926.6618 & 15.65 & 0.03 \\
58927.6389 & 15.78 & 0.03 \\ 
\bottomrule
\end{tabular}
\end{table}

\begin{table}[!t]
\centering
\caption{V-band photometry of 1ES 0647+250 from VBO.Here, $V$ and $\sigma_V$ are in magnitude units.}
\label{table-3}
\small
\begin{tabular}{ccc}
\toprule
MJD & V & $\sigma_V$ \\
\midrule
58874.7910 & 15.76 & 0.04 \\
58875.6722 & 15.68 & 0.03 \\
58876.6799 & 15.75 & 0.03 \\
58877.6785 & 15.74 & 0.03 \\
58877.7847 & 15.77 & 0.03 \\
58878.6160 & 15.75 & 0.04 \\
58878.6958 & 15.76 & 0.03 \\
58878.7986 & 15.80 & 0.04 \\
58879.6110 & 15.74 & 0.06 \\
58879.6910 & 15.73 & 0.05 \\
58880.6236 & 15.72 & 0.06 \\
58880.7118 & 15.88 & 0.05 \\
58881.6278 & 15.69 & 0.06 \\
58881.6813 & 15.41 & 0.05 \\
58894.5944 & 15.58 & 0.03 \\
58894.6944 & 15.59 & 0.03 \\
58895.5972 & 15.63 & 0.03 \\
58895.7076 & 15.68 & 0.03 \\
58896.5917 & 15.57 & 0.03 \\
58897.6042 & 15.59 & 0.03 \\
58897.7076 & 15.64 & 0.03 \\
58898.5889 & 15.60 & 0.03 \\
58898.6917 & 15.61 & 0.04 \\
58898.7917 & 15.58 & 0.05 \\
58899.5910 & 15.65 & 0.03 \\
58899.6938 & 15.68 & 0.03 \\
58899.7778 & 15.59 & 0.04 \\
58900.5875 & 15.53 & 0.04 \\
58900.6896 & 15.66 & 0.04 \\
58901.5868 & 15.65 & 0.04 \\
58904.6507 & 15.55 & 0.03 \\
58905.7486 & 15.59 & 0.04 \\
58905.7857 & 15.63 & 0.05 \\
58906.7635 & 15.74 & 0.05 \\
58907.7611 & 15.73 & 0.05 \\
58924.6097 & 15.65 & 0.03 \\
58925.6096 & 15.71 & 0.03 \\
58925.6931 & 15.64 & 0.03 \\
\bottomrule
\end{tabular}
\end{table}

\section{Data Reduction}
The analysis of the 2020 flaring episode of 1ES~0647+250 requires a cohesive dataset across multiple filters and epochs. To achieve this, we performed independent data reduction for our VBO observations and integrated it with high-cadence ZTF survey data. The following subsections detail the instrumental corrections, aperture photometry, and standard magnitude transformations used to minimize systematic errors across the light curves.
\subsection{\textbf{VBO Data Processing}}
The raw CCD frames were processed using the Image Reduction and Analysis Facility package ({\it IRAF}; \citealt{1986SPIE..627..733T}) to carry out the image reduction following standard procedures, that involved bias subtraction, flat field correction and cosmic ray removal. These bias frames were acquired at regular intervals during the night and flat field frames were taken at dawn and/or dusk. We carried out aperture photometry of the blazar and the comparison stars using an aperture radius of 2 $\times$ FWHM. We arrived at this optimum aperture by carrying out photometry for a sequence of aperture radii and using that aperture where the signal-to-noise ratio is maximum. 

The derived instrumental magnitudes of the blazar were brought to the standard system using the technique of differential photometry. In this method, the brightness of the blazar was measured relative to several non-variable comparison stars within the same FoV. This approach significantly reduces the effects of atmospheric transparency variations and instrumental fluctuations. The stability of the comparison stars was verified by examining the differential magnitudes between the selected standard stars, ensuring that no significant variability was present during the observing period. The photometric uncertainties were estimated by propagating the instrumental measurement errors obtained during the aperture photometry process, thereby providing reliable error estimates for the derived blazar magnitudes. For differential photometry we used four stars in the field of the blazar namely  A, C, F and H. The observed field of 1ES~0647+250 with the blazar and the comparison stars marked is given in Figure \ref{figure-1}. By maintaining a consistent set of comparison stars across all 107 epochs, we ensured a high degree of photometric precision throughout the 2020 flaring episode. The standard magnitudes of those stars in $B$,$V$, and $R$ filters were obtained from the online database\footnote{ Photometric sequence for 1ES~0647+250 available at \url{http://quasar.square7.ch/fqm/0647+250.html}}. 

The final dataset includes Modified Julian Date (MJD) values along with the corresponding magnitudes and measurement uncertainties ($\sigma$) in each filter. The calibrated photometric measurements of 1ES~0647+250 in the $R$, $V$, and $B$ bands are presented in Tables~\ref{table-2},~\ref{table-3}, and~\ref{table-4}, respectively. The typical photometric uncertainties range between 0.03 and 0.06 magnitudes. The optical variability of the source in the $B$, $V$, and $R$ filters plotted as magnitude versus time (MJD) is shown Figure \ref{figure-2}.

\subsection{\textbf{ZTF Data Processing}}
Publicly available data for 1ES~0647+250 was retrieved from the ZTF Science Data System (ZSDS). The ZTF light curves in the $z_g$ and $z_r$ bands were already processed through the standard ZTF pipeline, which includes bias subtraction, flat-fielding, and PSF (Point Spread Function) photometry (\citealt{2019EPSC...13.1163M}).

We applied an additional quality filtering criterion to the retrieved data by selecting only observations flagged as \texttt{"clear"} thereby ensuring high photometric accuracy. To enable a direct comparison with our VBO $R$-band data, the ZTF magnitudes were converted from the $z_{\text{AB}}$ photometric system to the standard Johnson--Cousins system wherever necessary. This procedure ensures uniformity in the photometric scale and minimizes possible systematic offsets between different datasets.The adoption of this combined approach allowed us to extend the temporal baseline of our analysis and to independently confirm the onset of the flare at $\text{MJD}~58853.44$.

\begin{table}[!t]
\centering
\caption{B-band photometry of 1ES 0647+250 from VBO.Here, $B$ and $\sigma_B$ are in magnitude units.}
\label{table-4}
\small
\begin{tabular}{ccc}
\toprule
MJD & B & $\sigma_B$ \\
\midrule
58876.6278 & 15.50 & 0.05 \\
58877.6271 & 15.50 & 0.05 \\
58878.7285 & 15.53 & 0.05 \\
58878.7618 & 15.59 & 0.06 \\
58894.6250 & 15.43 & 0.06 \\
58894.7542 & 15.40 & 0.04 \\
58895.6271 & 15.39 & 0.05 \\
58895.6708 & 15.48 & 0.04 \\
58896.6215 & 15.36 & 0.05 \\
58896.6563 & 15.45 & 0.04 \\
58897.6722 & 15.44 & 0.05 \\
58897.7569 & 15.36 & 0.05 \\
58898.6194 & 15.35 & 0.05 \\
58898.6569 & 15.36 & 0.05 \\
58899.6229 & 15.49 & 0.04 \\
58900.6194 & 15.53 & 0.05 \\
58904.6826 & 15.42 & 0.04 \\
58904.7243 & 15.29 & 0.05 \\
58924.6625 & 15.43 & 0.05 \\
58925.6576 & 15.50 & 0.05 \\
58926.5965 & 15.45 & 0.04 \\
58926.6792 & 15.41 & 0.04 \\
58927.5743 & 15.51 & 0.06 \\
\bottomrule
\end{tabular}
\end{table}
\subsection{\textbf{Combined Data Set}}
To provide continuous temporal coverage of the 2020 flare, we integrated our ground-based $B, V, R$ photometry from VBO with public-domain $z_g$ and $z_r$ data from ZTF. 

Because these observations originate from different telescope systems and filter sets, we carefully cross-matched the datasets during overlapping periods.Constant offsets of -0.21 mag and +0.41 mag were applied to the VBO $R$- and $B$- band data, respectively, to align them with the ZTF $r$- and $g$- band observations shown in Figure~\ref{figure-2}. Although the filters are not identical, the ZTF $r$- and VBO $R$- bands exhibit similar variability behavior after offset correction, indicating consistent photometric trends between the datasets. The uncertainties in the VBO data are relatively larger than those of the ZTF data due to differences in instrumentation, observing conditions, and data reduction procedures. To ensure a uniform analysis, all magnitudes were corrected for Galactic extinction using the values from \citet {2011ApJ...737..103S}, assuming an $R_V = 3.1$ reddening law. This unified dataset enables a robust cross-correlation analysis and a detailed study of the flare's rise and decay phases.

The combined light curve illustrates the multi-band evolution of 1ES~0647+250 over a period of approximately 24~days, revealing several key features of its 2020 activity. The onset of the flare was initially detected in the ZTF $z_g$ data, which showed the beginning of increased activity around $\text{MJD} \approx 58853.44$ at a magnitude of $15.91$ mag. This was followed by intensive monitoring from the VBO, starting at $\text{MJD} \approx 58874.77$, which captured the source during its peak activity. Notably, the $R$-band magnitude reached its brightest state of $15.74$ mag at $\text{MJD} \approx 58876.66$.

We found good agreement in the overall variability morphology between the ZTF and VBO observations, particularly between the ZTF $z_r$ and VBO $R$ filters, which overlap in wavelength and exhibit comparable flux variations within the range of $\sim15.7$--$15.8$ magnitudes. Furthermore, the close synchronization across all five filters ($B, V, R, z_g, z_r$) suggests that the optical emission is predominantly governed by a single synchrotron component originating from the relativistic jet.

The photometric data for 1ES~0647+250, presented in Tables~\ref{table-2}, ~\ref{table-3}, and~\ref{table-4}, include the MJD, the observed magnitude in each filter, and the corresponding $1\sigma$ uncertainties. Since the observations in the $R$, $V$, and $B$ bands were performed sequentially, each measurement is listed with its specific MJD to accurately reflect the non-simultaneous nature of the multi-band monitoring.

\section{Results}
We combined the optical light curves constructed from VBO and ZTF observations to obtain a comprehensive view of the variability behavior of the blazar during the 2020 flaring episode. The ZTF data cover a longer time span, while the VBO observations provide higher-cadence measurements over a specific interval of the flare. The VBO observations span the period from MJD $\approx 58874$ to $58894$, capturing a particular segment of the broader flaring activity recorded by ZTF, which extends from approximately MJD $\approx 58194$ (March 17, 2018) onwards. 

Thus, the VBO data represent a focused monitoring window within the broader variability trend detected by ZTF. A comparison between the ZTF $z_r$- band and the VBO $R$- band shows consistent magnitude levels, typically in the range of $15.7$--$15.8$ mag (Table~\ref{table-5}). This agreement confirms the reliability of the photometric calibration and demonstrates strong consistency between the two independent observational datasets.

The ZTF light curve shows relatively steady brightness during MJD $58853$--$58865$, followed by enhanced variability. The VBO observations capture this later stage of the flare, revealing noticeable short-term flux fluctuations across the $B$, $V$, and $R$-bands.

\begin{table}[htbp]
\centering
\caption{A representative sample of the optical photometry of 1ES~0647+250 during the 2020 flaring period. The full dataset includes $BVR$ filters from VBO and $z_g, z_r$ filters from ZTF. Magnitudes are as observed, and errors represent the $1\sigma$ uncertainty.}
\label{table-5}
\begin{tabular}{ccccc}
\toprule
MJD & Filter & Magnitude & Error & Telescope \\
\midrule
58853.44 & $z_g$ & 15.91 & 0.01 & ZTF \\
58874.77 & $R$    & 15.79  & 0.03  & VBO \\
58874.79 & $V$    & 15.76  & 0.04  & VBO \\
\bottomrule
\end{tabular}
\end{table}

\subsection{\textbf{Variability Amplitude Analysis}}

\begin{figure}[!t]
\centering
\includegraphics[width=\columnwidth]{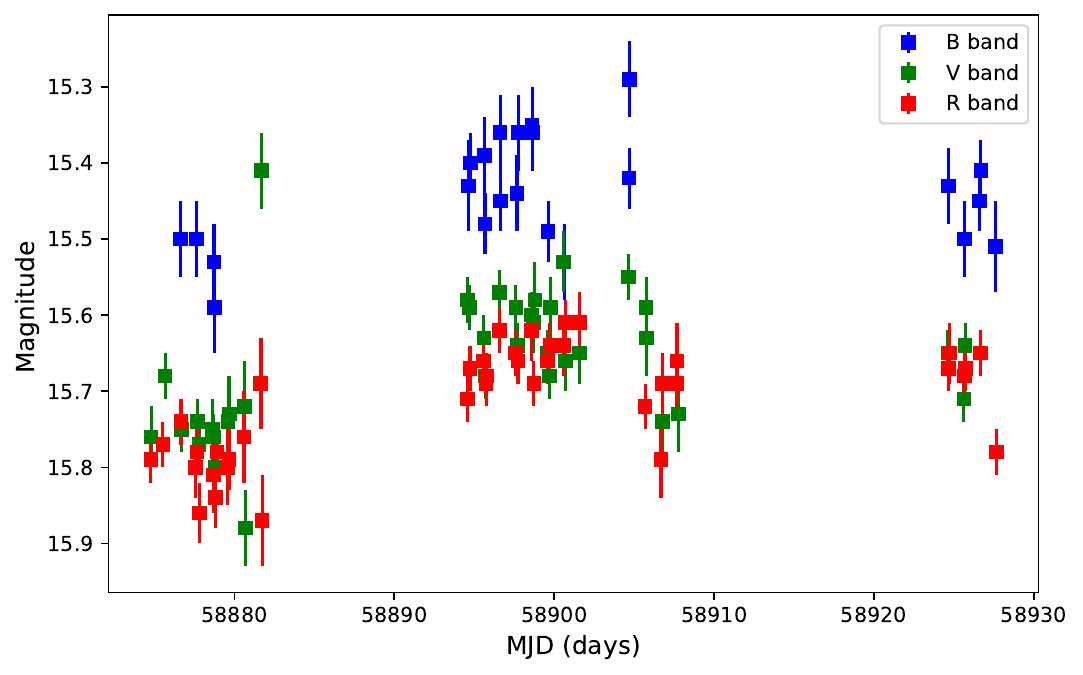}
\caption{Light curves of the source 1ES 0647+250 in $B, V,$ and $R$ filters obtained from VBO.}
\label{figure-3}
\end{figure}

The light curves of the blazar 1ES 0647+250 in the $B, V,$ and $R$ filters are shown in Figure~\ref{figure-3}. To characterize the intrinsic optical variability during the 2020 flare state, we estimated the intrinsic variability amplitude ($\sigma_m$). Following \citet{2007AJ....134.2236S}, $\sigma_m$ was calculated by subtracting the contribution of photometric uncertainties from the observed variance of the light curves:
\begin{equation}
\sigma_m = \sqrt{S^2 - \langle \sigma_{\mathrm{err}}^2 \rangle}
\end{equation}
where $S^2$ is the observed variance of the light curve and $\langle \sigma_{\mathrm{err}}^2 \rangle$ is the mean square of the individual measurement errors ($\sigma_i$). These are defined as:
\begin{equation}
S^2 = \frac{1}{n-1} \sum_{i=1}^{n} (m_i - \bar{m})^2
\end{equation}
and
\begin{equation}
\langle \sigma_{\mathrm{err}}^2 \rangle = \frac{1}{n} \sum_{i=1}^n \sigma_i^2
\end{equation}
where $n$ is the number of observed data points in the light curve and $\bar{m}$ is their weighted average. We considered $\sigma_m$ to be a real value if $S^2 > \langle \sigma_{\mathrm{err}}^2 \rangle$; otherwise, $\sigma_m$ was set to zero.

The derived variability amplitudes indicate significant intrinsic variability in all three bands during the 2020 flare state. The $B$-band exhibits the largest variability amplitude, followed by the $V$- and $R$-bands, as summarized in Table~\ref{table-6}. In all filters, the observed variance exceeds the mean squared photometric uncertainty ($S^{2} > \langle \sigma_{\mathrm{err}}^{2} \rangle$), confirming the presence of genuine intrinsic variability. The $R$-band light curve contains the largest number of observations ($N = 40$), whereas the $B$-band has relatively fewer measurements ($N = 23$). Overall, the observed variability over the $\sim$30-day interval (MJD 58874--58904) is consistent with the stochastic optical variability commonly observed in BL Lac objects during active states.

\begin{figure*}[!t]
\centering
\includegraphics[width=\textwidth]{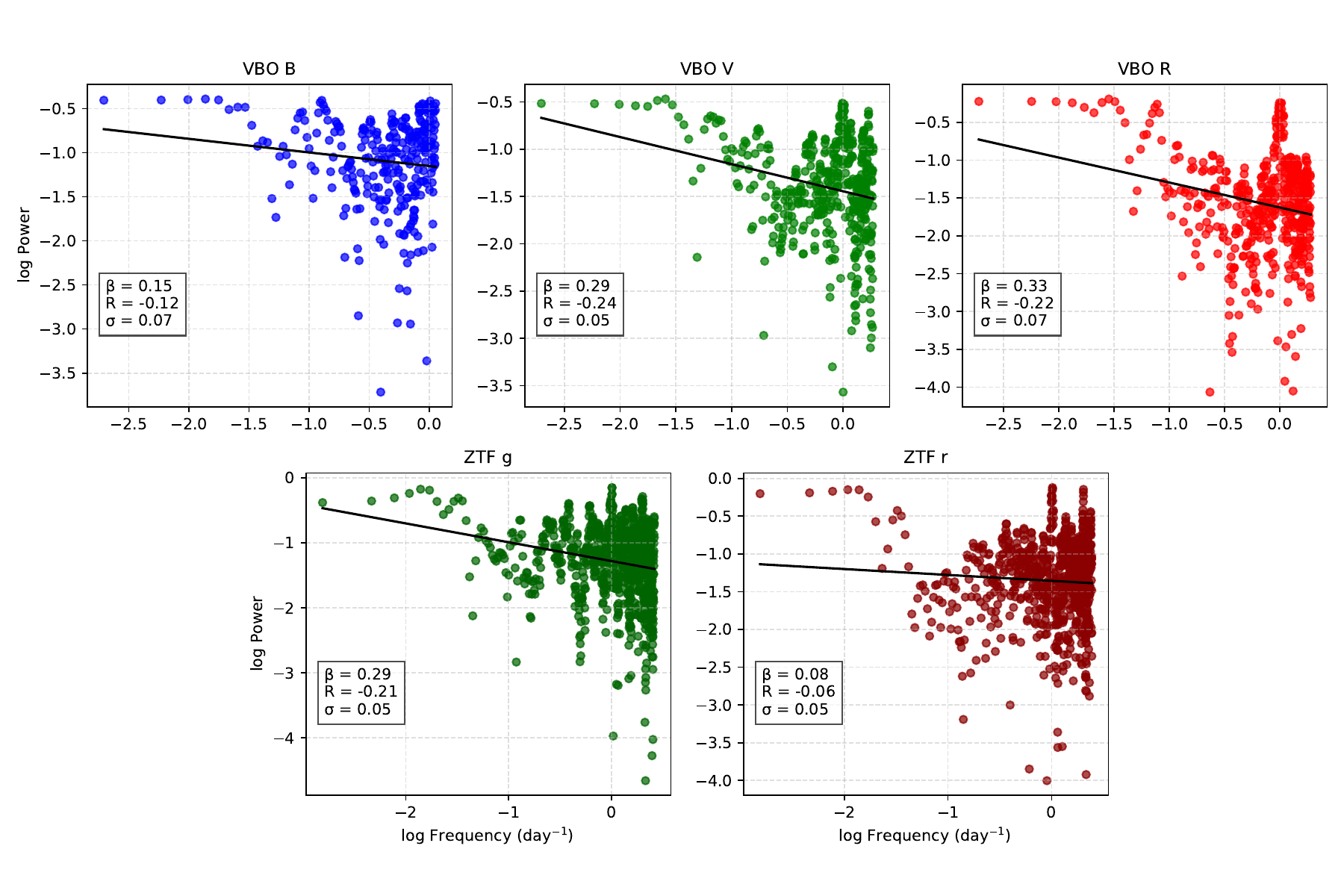}
\caption{Power Spectral Density plots of 1ES~0647+250 for the VBO ($B, V, R$) and ZTF ($g, r$) bands. The raw, unbinned Lomb-Scargle periodograms are shown to preserve frequency resolution. The best-fit power-law slopes ($\beta$), correlation coefficients ($R$), and noise levels ($\sigma$) are indicated for each band. The consistency between the VBO $V$-band and ZTF $g$-band slopes ($\beta = 0.29$) highlights the reliability of the variability detection.}
\label{figure-4}
\end{figure*}

\begin{table}[htbp]
\centering
\caption{Variability amplitude ($\sigma_{m}$) of the BL Lac object 1ES 0647+250 in the B, V and R optical bands derived from VBO observations.}
\label{table-6}

\small

\begin{tabular}{lcccc}
\toprule
Band & $N$ & Mean Magnitude & $\sigma_{m}$ \\
\midrule
$R$ & 40 & 15.71 & 0.036  \\
$V$ & 38 & 15.66 & 0.039  \\
$B$ & 23 & 15.44 & 0.048 \\
\bottomrule
\end{tabular}

\end{table}

\subsection{\textbf{Power Spectral Density Analysis}}

To investigate the variability characteristics of 1ES~0647+250 in the frequency domain, we performed a power spectral density (PSD) analysis of the optical light curves. The PSD plots are as shown in Figure~\ref{figure-4}. To account for the uneven sampling of the observations, we employed the Lomb--Scargle periodogram (LSP; \citealt{1976Ap&SS..39..447L,1982ApJ...263..835S}). The LSP was computed over a frequency range extending from
$f_{\min} = \frac{1}{T}$
to the pseudo-Nyquist frequency
$f_{\max} = \left\langle \frac{1}{2\Delta t} \right\rangle$,
where $T$ is the total observational duration and $\Delta t$ is the mean sampling interval.

Given the relatively limited temporal coverage and modest number of observations during the 2020 flare, logarithmic binning of the periodogram was not applied, as such binning could substantially reduce the available frequency information. Instead, the unbinned PSD was retained to preserve the maximum frequency resolution and to illustrate the scatter at high temporal frequencies, which likely reflects the contribution of measurement uncertainties. The power-law spectral index ($\beta$) was estimated by fitting a linear model to the log--log periodogram, assuming a power-law form:

\begin{equation}
    P(f) \propto f^{-\beta}
\end{equation}

where $\beta$ is the spectral index \citep{2002MNRAS.332..231U}. The derived PSD parameters are summarized in Table~\ref{table-7}.

The VBO data show relatively low spectral indices ($\beta < 1$), with the $R$- band exhibiting the most structured variability ($\beta = 0.33$). While long-term blazar variability studies often report that the slope ($\beta$) indicates the type of noise present in the light curve, with $\beta \approx 1$--$2$ typically representing stochastic processes commonly observed in blazars \citep{1978ComAp...7..103P}, the shallower slopes obtained here suggest that, over the short timescales probed in this study, the variability is closer to flicker- or white-noise behavior. The agreement between the VBO $V$- band and ZTF $g$- band ($\beta = 0.29$) across independent instruments further supports the consistency of the observed variability characteristics.

The relatively low noise level ($\sigma \approx 0.05$--$0.07$) across both datasets indicates that the observed variability, although modest in amplitude, is intrinsic to the source. Overall, the PSD results suggest that 1ES~0647+250 exhibits stochastic optical variability, plausibly associated with turbulent processes and/or localized particle acceleration mechanisms, such as magnetic reconnection, within the relativistic jet. Furthermore, the absence of a steep PSD slope ($\beta \gtrsim 1$) suggests that no strong long-timescale correlated variability dominates the observed optical emission during the flare episode. Similar shallow PSD behaviour has been reported in short-duration optical monitoring studies of blazars, where rapid stochastic fluctuations are often attributed to localized turbulence and inhomogeneities within the jet environment. However, given the limited temporal coverage and uneven sampling of the optical light curves, the inferred PSD slopes should be interpreted with caution. 

Nevertheless, the broadly consistent PSD behaviour across the VBO and ZTF datasets suggests that the observed optical variability pattern is robust against instrumental differences. Future high-cadence and longer-duration optical monitoring would be valuable to better constrain the PSD slope and to investigate the possible presence of characteristic variability timescales in 1ES~0647+250.

\begin{table}[htbp]
\centering
\caption{Power spectral density analysis parameters}
\label{table-7}
\small
\begin{tabular}{lccc}
\toprule
Filter & Spectral Index  & Correlation  & Noise Level  \\
       &  ($\beta$)      & ($R$)        & ($\sigma$)   \\ \midrule
VBO $B$ & 0.15 & $-0.12$ & 0.07 \\
VBO $V$ & 0.29 & $-0.24$ & 0.05 \\
VBO $R$ & 0.33 & $-0.22$ & 0.07 \\
ZTF $g$ & 0.29 & $-0.21$ & 0.05 \\
ZTF $r$ & 0.08 & $-0.06$ & 0.05 \\
\bottomrule
\end{tabular}
\end{table}

\subsection{\textbf{ Spectral Variability Analysis}}

In addition to flux variability, blazars are known to exhibit spectral variability, which can provide important insights into the underlying emission mechanisms and particle acceleration processes within relativistic jets \citep{2012ApJ...756...13B,2022MNRAS.516.5376N}. To investigate the spectral behaviour of 1ES~0647+250 during the 2020 flare, we constructed color--magnitude diagrams using quasi-simultaneous observations from VBO and ZTF. Such analyses are commonly used to examine possible spectral trends, including bluer-when-brighter (BWB) or redder-when-brighter (RWB) behaviour.The derived slopes ($m$), Pearson correlation coefficients ($R$), and residual dispersions ($\sigma$) are summarized below.

\subsubsection{\textbf{Color--Magnitude Behaviour}}

\begin{figure*}[!t]
\centering
\includegraphics[width=0.96\textwidth]{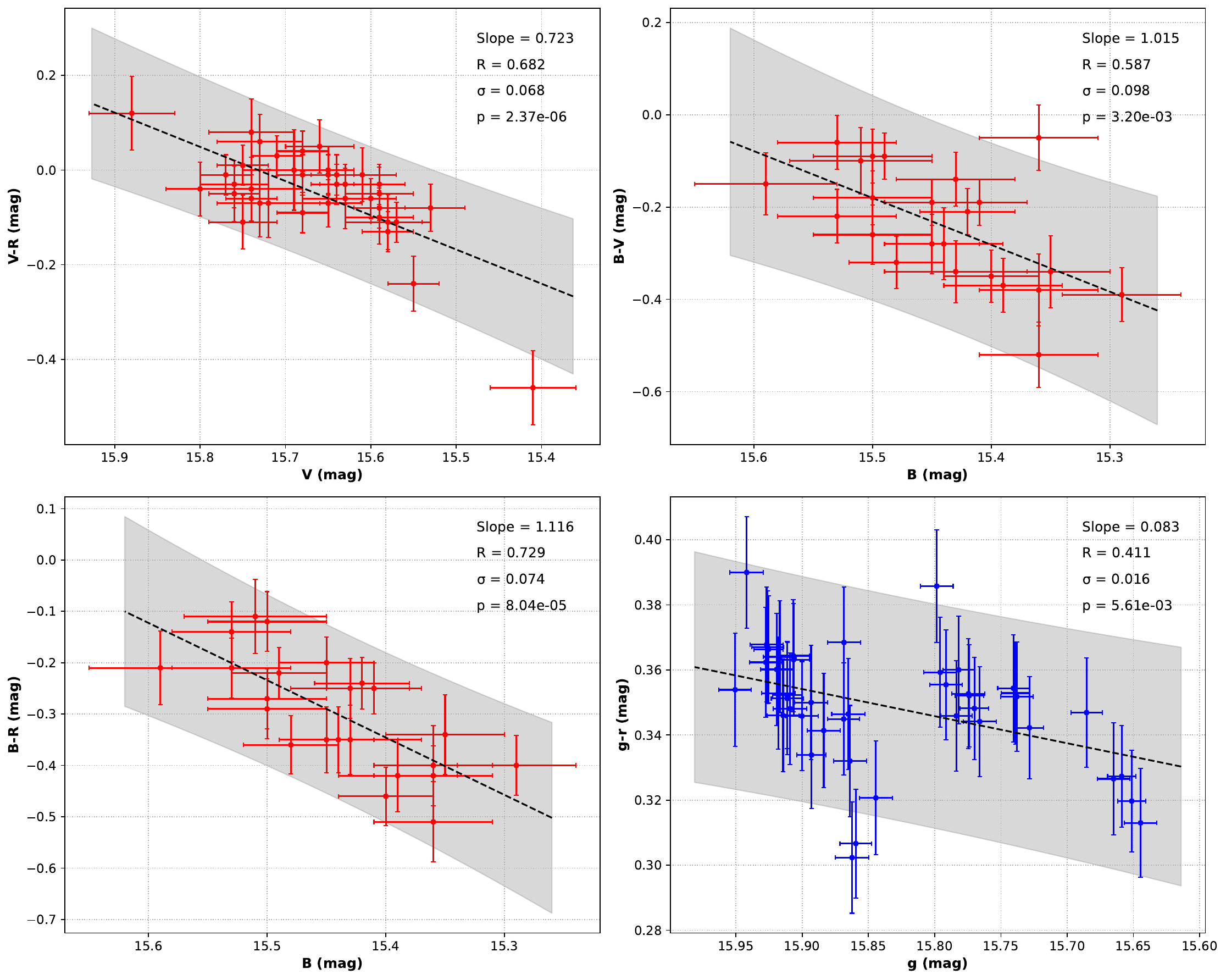}
\caption{Optical color--magnitude diagrams of 1ES~0647+250 from the VBO and ZTF observations. The panels show the $(V-R)$ versus $V$, $(B-V)$ versus $B$, $(B-R)$ versus $B$, and $(g-r)$ versus $g$ relations. Solid lines show the linear regression fits, and shaded regions indicate the 95\% prediction intervals.}
\label{figure-5}
\end{figure*}

The ($V-R$) versus $V$ colour magnitude diagram (CMD) shown in Figure~\ref{figure-5}, clearly indicates hardening as the source brightens. We also constructed ($B-V$) versus $B$, $(B-R)$ versus $B$ and ($g-r$) versus $g$, CMDs. All CMDs (Figure~\ref{figure-5}) reveal statistically significant chromatic variability, consistent with a BWB trend.

Such chromatic behaviour is commonly observed in HSP BL Lac objects, where enhanced synchrotron emission from freshly accelerated relativistic electrons produces larger variability amplitudes at shorter wavelengths \citep{1997A&A...327...61G,2004A&A.419..25F}. Within the shock-in-jet framework \citep{1985ApJ...298..114M}, shocks propagating along the relativistic jet accelerate particles to higher energies, naturally leading to wavelength-dependent variability and the observed BWB behaviour.

The observed BWB trend in 1ES~0647+250 is therefore consistent with its classification as a HSP object and supports a synchrotron-dominated emission scenario. Similar behaviour has been widely reported in BL Lac objects, whereas Flat Spectrum Radio Quasars (FSRQs) often exhibit a redder-when-brighter trend \citep[][and references therein]{2003MNRAS.339.1237V,2022MNRAS.515.1952A,2023MNRAS.522..111A}.

\subsubsection{\textbf{Statistical Significance of the Color--Magnitude Trends}}
To assess the reliability of the derived linear fits, we calculated the 95\% confidence intervals for each regression. The confidence interval, shown as the shaded region in Figure~\ref{figure-5}, represents the range within which the true regression line is expected to lie with 95\% probability.

\begin{equation}
{\Delta y = t_{\alpha/2, n-2} \cdot s_{e} \sqrt{\frac{1}{n} + \frac{(x - \bar{x})^2}{\sum (x_i - \bar{x})^2}}}
\end{equation}

where $\Delta y$ is the uncertainty in the predicted color index, $s_{e}$ is the standard error of the residuals, $x$ is the observed magnitude, and $\bar{x}$ is the mean magnitude of the dataset. 

The characteristic ``bow-tie'' or curved shape of the shaded region arises from the $(x - \bar{x})^2$ term in the formula. This indicates that our model is most precise near the mean magnitude of the flare, with uncertainty naturally increasing toward the extreme bright and faint ends of the observed range. This rigorous approach ensures that the BWB trends reported are statistically significant across the entire observed flaring period.

The statistical significance of the color--magnitude correlations was further evaluated using the Pearson correlation coefficient ($R$) and the associated $p$-values. All correlations derived from the VBO dataset are highly significant ($p < 10^{-4}$), indicating that the observed BWB trends are statistically robust and unlikely to arise from random fluctuations.

\subsection{\textbf{Cross-Correlation Analysis}}

We determined the inter-band time lag between the $V$- and $R$- band light curves using the Interpolated Cross-Correlation Function (ICCF; \citealt{1986ApJ...305..175G}, \citealt{1987ApJS...65....1G}). The lag was estimated from the centroid ($\tau_{\rm cent}$) of the cross-correlation function (CCF), computed using all points with correlation coefficients greater than 80\% of the peak value ($r_{\rm max}$; \citealt{1998PASP..110..660P}). 

\begin{figure}
\hspace*{-0.1cm}\includegraphics[scale=0.55]{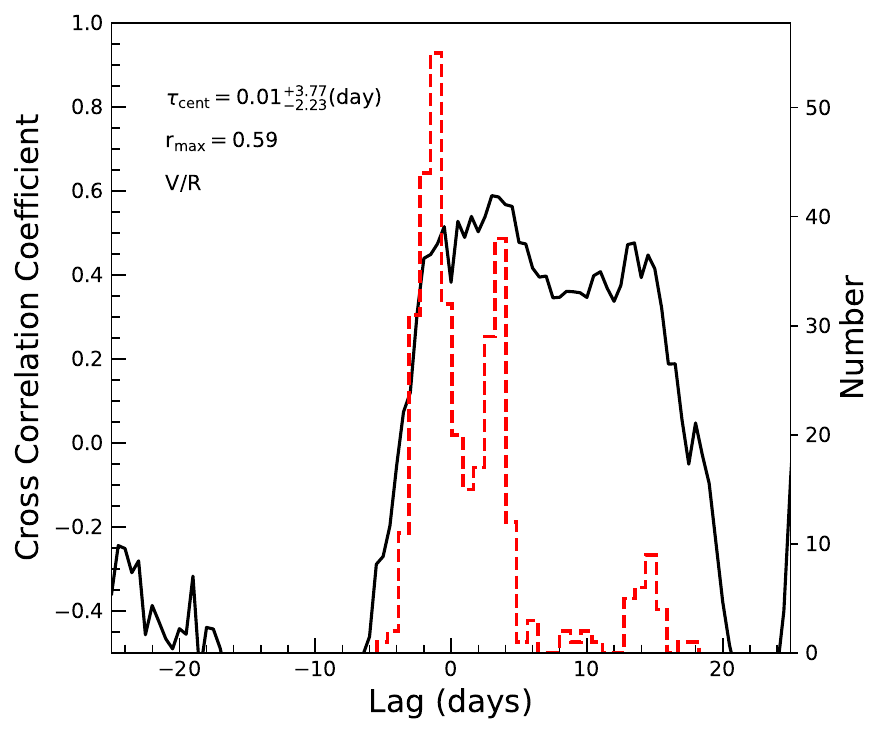}
    \caption{Cross-correlation between $V$- and $R$- band light curves of 1ES~0647+250. The black solid curve represents the ICCF with a maximum correlation coefficient of r$_{max}$ = 0.59. The dotted red histogram represents the distribution of the lag determined over 100 Monte Carlo simulations, with the centroid lag $\tau_{cent}$ indicated.(see Section 4.4).}
\label{figure-6}
\end{figure}

The uncertainty in the measured lag was estimated using a model-independent Monte Carlo technique based on flux randomization and random subset selection (FR/RSS; \citealt{1998PASP..110..660P,1999ApJ...526..579W,2004ApJ...613..682P}).

As shown in Figure~\ref{figure-6}, we obtained a lag of $\tau_{\rm cent} = 0.01^{+3.77}_{-2.23}$ days, indicating no significant delay between the $V$ and $R$ bands within the uncertainties. The absence of a measurable time lag suggests that the optical emission in these bands likely originates from the same emission region within the relativistic jet.These findings are consistent with a co-spatial origin of the optical emission and support a scenario in which rapid particle acceleration and synchrotron cooling operate within a compact region of the jet.

\subsection{\textbf{Structure Function Analysis}}

To investigate the variability behavior of 1ES~0647+250 over different timescales, we performed a first-order structure function (SF) analysis using both the VBO and ZTF datasets. The SF is a powerful statistical tool for characterizing variability amplitudes as a function of time lag and is particularly suitable for unevenly sampled astronomical light curves. The first-order SF, widely used to characterize variability in unevenly sampled astronomical time series \citep{1985ApJ...296...46S}, is defined as

\begin{equation}
SF(\tau)=\sqrt{
\frac{1}{N(\tau)}
\sum [m(t)-m(t+\tau)]^2
-\sigma_{\rm noise}^{2}
}
\end{equation}

where $m(t)$ is the observed magnitude at time $t$, $\tau$ is the time lag, $N(\tau)$ is the number of data pairs contributing to a given lag interval, and $\sigma_{\rm noise}^{2}$ represents the contribution from photometric measurement uncertainties.
\begin{figure*}[t]
\centering
\includegraphics[width=\textwidth]{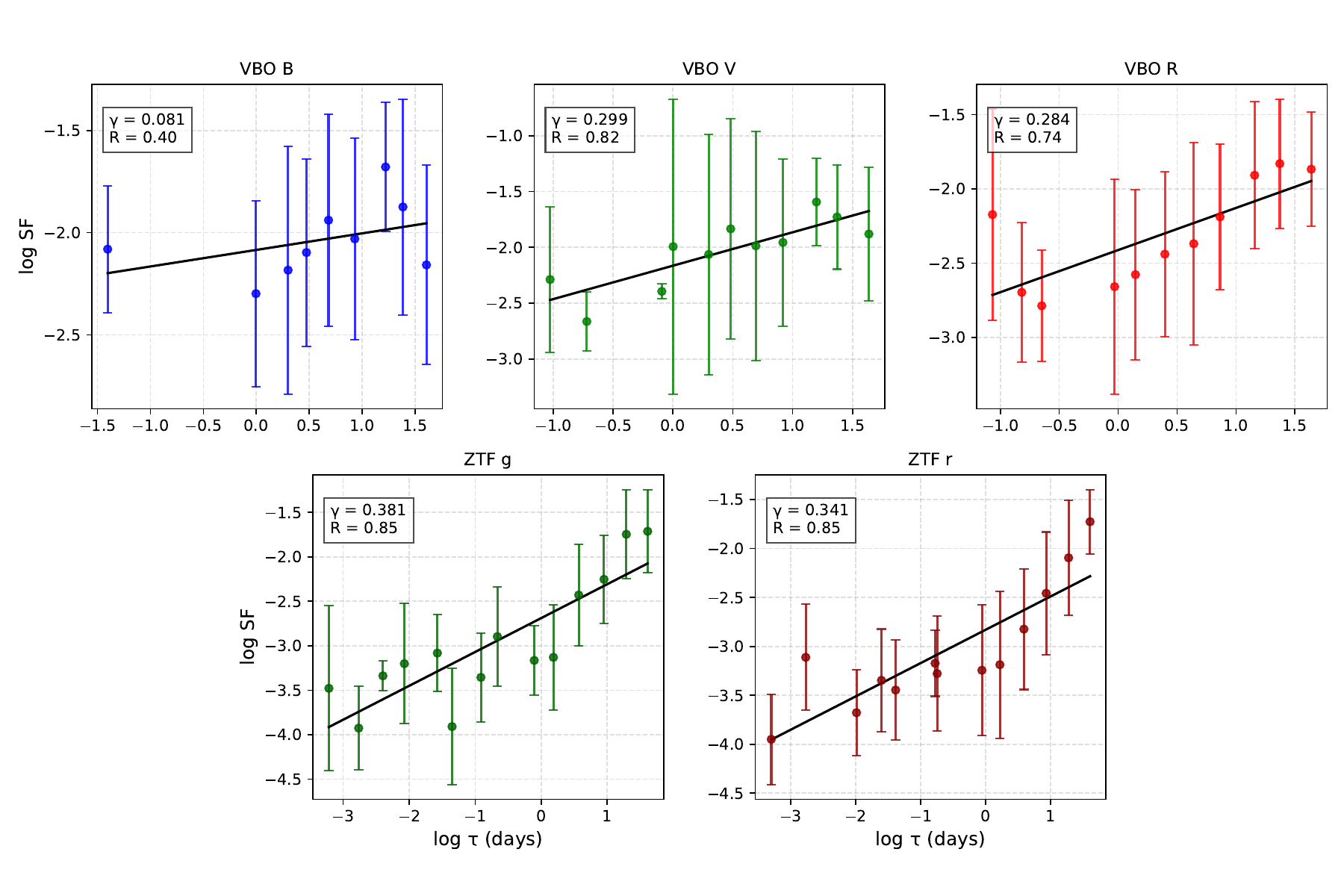}
\caption {First-order SF plots of 1ES~0647+250 in the VBO ($B$, $V$, $R$) and ZTF ($g$, $r$) bands. The SFs were computed using logarithmically spaced discrete time-lag bins. The data points represent the mean SF amplitudes within each lag bin, and the error bars denote the corresponding standard deviations. The solid black lines show the best-fit linear regressions in the log--log plane. The derived slopes ($\gamma$) and Pearson correlation coefficients ($R$) are indicated in each panel.}
\label{figure-7}
\end{figure*}
Following the standard approach adopted in previous variability studies \citep{2004ApJ...601..692V,2011PASP..123..634W,2016MNRAS.459.2787K,2020Ap&SS.365..182A}, the SF was computed using logarithmically spaced discrete time-lag bins. For each lag interval, the representative SF value was obtained by averaging all contributing data pairs within the corresponding bin, while the associated standard deviation was adopted as the uncertainty. This binning procedure reduces statistical scatter and provides a more robust characterization of variability trends across different timescales.

Figure~\ref{figure-7} presents the SF behavior derived from both datasets. The VBO observations exhibit comparatively shallow SF slopes, with $\gamma = 0.081$, $0.299$, and $0.284$ for the $B$, $V$, and $R$ bands, respectively. In particular, the weak correlation observed in the $B$ band ($R = 0.40$) suggests a less coherent variability trend over the sampled timescales. In contrast, the ZTF archival data shows comparatively steeper slopes of $\gamma = 0.381$ and $0.341$ for the $g$ and $r$ bands, respectively, indicating stronger long-term variability compared to the shorter-timescale variability captured by the high-cadence VBO observations.

The VBO dataset primarily probes short-timescale variability associated with the 2020 flare activity, whereas the broader temporal coverage of the ZTF archival data enables the characterization of longer-term stochastic variability. The observed power-law behavior of the SF is consistent with stochastic synchrotron variability commonly observed in blazar jets \citep{1996A&A...305...42H,2011A&A...536A..84V,2022ApJ...933...42A}.

The Pearson correlation coefficients of the linear fits are relatively weak for the VBO data and moderately stronger for the ZTF data, indicating that the long-term variability follows a more coherent power-law trend than the short-term flare variability. The comparatively shallow SF slopes in the VBO observations may arise from rapid micro-variability, localized turbulence, or measurement uncertainties during the flare state.

\section{Discussion}

The optical variability observed in the blazar 1ES~0647+250 during the 2020 flare provides important insights into the physical processes operating within its relativistic jet. The combined VBO and ZTF observations reveal significant short-term variability across all optical bands, supported by improved temporal coverage during the active state.

The variability analysis indicates moderate intrinsic variability, with the largest amplitude observed in the $B$-band. The detection of variability in all optical bands suggests that the observed emission is predominantly governed by non-thermal synchrotron radiation produced by relativistic electrons within the jet.

The color--magnitude analysis reveals a statistically significant BWB trend, indicating spectral hardening during brighter states. Such behaviour is commonly observed in HSP blazars and is generally interpreted as enhanced synchrotron emission from freshly accelerated high-energy electrons. The confidence intervals exhibit a characteristic ``bow-tie'' structure, reflecting that the linear regression model is most strongly constrained near the mean brightness state of the flare, while uncertainties increase toward the extreme bright and faint states due to limited sampling. The observed spectral evolution is consistent with a shift of the synchrotron peak toward higher frequencies during the flare state.

The strong correlation between the $V$- and $R$-band light curves, together with the near-zero time lag, suggest that the optical emission originates from closely related or co-spatial emitting regions within the jet. The PSD analysis further indicates the presence of stochastic variability processes operating on short timescales. However, given the limited temporal coverage and uneven sampling of the optical light curves, the inferred PSD slopes should be interpreted with caution.

The observed variability characteristics can be interpreted within the framework of the standard shock-in-jet model \citep{1985ApJ...298..114M}, in which shocks propagating along the relativistic jet accelerate particles through first-order Fermi processes, leading to enhanced synchrotron emission during flaring states. The observed BWB trend, correlated multi-band variability, and spectral hardening are broadly consistent with enhanced synchrotron emission associated with freshly accelerated high-energy electrons.

Beyond diffusive shock acceleration, alternative particle acceleration and jet dissipation mechanisms may also play a significant role in driving the observed optical variability. Magnetic reconnection events within relativistic and highly magnetized jet regions can efficiently energize particles and trigger localized energy dissipation, thereby producing rapid flux variations and spectral evolution \citep{2013MNRAS.431..355G,2020NatCo..11.4176M}. Similarly, kink-instability-driven processes may facilitate magnetic energy release, turbulent particle acceleration, and enhanced synchrotron radiation \citep{2023MNRAS.521L..53A}. Consequently, the observed spectral hardening and bluer-when-brighter behaviour of 1ES~0647+250 during its active state may plausibly arise, at least in part, from the combined influence of these alternative jet dissipation mechanisms.

The optical variability properties observed in 1ES~0647+250 are broadly consistent with previous optical monitoring studies of $\gamma$-ray emitting blazars, where variability has been interpreted within the framework of synchrotron jet emission and leptonic scenarios \citep{2022JApA...43...48P}.

Overall, the variability properties of 1ES~0647+250 during the 2020 flare are consistent with synchrotron-dominated emission from a dynamically evolving relativistic jet, similar to the behaviour commonly observed in HSP BL Lac objects \citep{2013ApJ...764..141G,2003ApJ...590..123V}.

\section{Conclusions}

In this study, we performed a multi-band optical photometric investigation of the $\gamma$-ray emitting blazar 1ES~0647+250 during its prominent 2020 flaring episode. By combining observations from VBO with archival data from ZTF, we achieved improved temporal coverage, enabling a detailed investigation of the source's optical variability properties. The principal findings of this work are summarized as follows:

\begin{enumerate}
\item The source exhibited significant and correlated flux variability across all optical bands. The variability amplitude was highest in the $B$ band, followed by the $V$ and $R$ bands, consistent with synchrotron-dominated emission commonly observed in HSP blazars.

\item The color--magnitude analysis revealed a statistically significant BWB trend across multiple filter combinations. This suggests spectral hardening during brighter states, indicating enhanced contribution from higher-energy synchrotron-emitting electrons.

\item Cross-correlation analysis between the $V$ and $R$ bands yielded a significant correlation ($r_{\text{max}} = 0.59$) with a near-zero centroid lag ($\tau \approx 0$ days), implying that the optical emission originates from closely related or co-spatial regions within the relativistic jet.

\end{enumerate}
 Overall, our results suggest that the observed variability in 1ES~0647+250 is governed by stochastic processes and particle acceleration within a dynamically evolving relativistic jet. Future long-term, densely sampled multiwavelength observations will be crucial for better constraining the physical mechanisms responsible for the variability and spectral evolution of this source.

%%Appendix

\section*{Acknowledgements}
The authors gratefully acknowledge the observing facilities provided by VBO, operated by the Indian Institute of Astrophysics, Kavalur, India. We thank the observatory staff for their support during the observations and for maintaining the telescope and instrumentation used in this study. This work has also made use of public data from ZTF. The ZTF is supported by the National Science Foundation and aninternational collaboration of partners. The survey data used in this research were obtained through the publicly available ZTF data archive. K.S.P. acknowledges Mr. P. Anbazhagan, Mr. G. Selvakumar, Mr. Ramachandran, and the VBO staff for their assistance during our observing runs. We thank the anonymous referees for constructive comments and suggestions that helped improve the quality of this manuscript.

\vspace{-1em}

% === PASTE YOUR BIBLIOGRAPHY HERE ===

% ====================================

\begin{thebibliography}{99}
\expandafter\ifx\csname natexlab\endcsname\relax\def\natexlab#1{#1}\fi

\bibitem[Agarwal et al.(2022a)]{2022MNRAS.515.1952A} 
Agarwal, A., Maitra, C., Gupta, A. C., Sagar, R., et al. 2022a, MNRAS, 515, 1952

\bibitem[Agarwal et al.(2023a)]{2023MNRAS.522..111A} 
Agarwal, A., Gupta, A. C., Wiita, P. J., et al. 2023a, MNRAS, 522, 111

\bibitem[Agarwal et al.(2023b)]{2023MNRAS.521L..53A}
Agarwal, A., et al. 2023b, MNRAS, 521, L53

\bibitem[Agarwal et al.(2022b)]{2022ApJ...933...42A} 
Agarwal, A., Pandey, A., Özdönmez, A., et al. 2022b, ApJ, 933, 42

\bibitem[Ajello et al.(2022)]{2022ApJS..263...24A} 
Ajello, M., Angioni, R., Axelsson, M., et al. 2022, ApJS, 263, 24

\bibitem[Aleksi{\'c} et al.(2016)]{2016A&A...593A..98L} 
Aleksi{\'c}, J., Ansoldi, S., Antonelli, L. A., et al. 2016, A\&A, 593, A98

\bibitem[Aleksi{\'c} et al.(2011)]{2011ApJ...742...43A} 
Aleksi{\'c}, J., Antonelli, L. A., et al. 2011, ApJ, 742, 43

\bibitem[Angel \& Stockman(1980)]{1980ARA&A..18..321A} 
Angel, J. R. P., Stockman, H. S. 1980, ARA\&A, 18, 321

\bibitem[Anjum et al.(2020)]{2020Ap&SS.365..182A}
Anjum, A., Stalin, C.~S., Chand, H., \& Zhang, X.-G. 2020, Ap\&SS, 365, 182

\bibitem[Bellm et al.(2019)]{2019PASP.131.068003B} 
Bellm, E. C., Kulkarni, S. R., Graham, M. J., et al. 2019, PASP, 131, 068003

\bibitem[Bonning et al.(2012)]{2012ApJ...756...13B} 
Bonning, E. W., Urry, C. M., Bailyn, C., et al. 2012, ApJ, 756, 13

\bibitem[Costamante \& Ghisellini(2002)]{2002A&A.384.56C} 
Costamante, L., \& Ghisellini, G. 2002, A\&A, 384, 56

\bibitem[Fiorucci et al.(2004)]{2004A&A.419..25F} 
Fiorucci, M., Ciprini, S., Tosti, G. 2004, A\&A, 419, 25

\bibitem[Gaskell \& Peterson(1987)]{1987ApJS...65....1G} 
Gaskell, C. M., \& Peterson, B. M. 1987, ApJS, 65, 1

\bibitem[Gaskell \& Sparke(1986)]{1986ApJ...305..175G} 
Gaskell, C. M., \& Sparke, L. S. 1986, ApJ, 305, 175

\bibitem[Gaur et al.(2013)]{2013ApJ...764..141G}
Gaur, H., Stalin, C.~S., Pandey, A., et al.\ 2013, ApJ, 764, 141

\bibitem[Ghisellini et al.(1997)]{1997A&A...327...61G} 
Ghisellini, G., Villata, M., Raiteri, C. M., et al. 1997, A\&A, 327, 61

\bibitem[Giannios(2013)]{2013MNRAS.431..355G}
Giannios, D. 2013, MNRAS, 431, 355

\bibitem[Gupta et al.(2008b)]{2008AJ....136.2359G} 
Gupta, A. C., Cha, S.-M., Lee, S., et al. 2008b, AJ, 136, 2359

\bibitem[Gupta et al.(2008a)]{2008AJ....135.1384G} 
Gupta, A. C., Fan, J. H., Bai, J. M., \& Wagner, S. J. 2008a, AJ, 135, 1384

\bibitem[Heidt \& Wagner(1996)]{1996A&A...305...42H} 
Heidt, J., \& Wagner, S. J. 1996, A\&A, 305, 42

\bibitem[Kapanadze(2009)]{2009MNRAS.398..832K} 
Kapanadze, B. Z. 2009, MNRAS, 398, 832

\bibitem[Kotilainen et al.(2011)]{2011A&A...534L...2K} 
Kotilainen, J. K., Hyvönen, T., Falomo, R., et al. 2011, A\&A, 534, L2

\bibitem[Koz{\l}owski(2016)]{2016MNRAS.459.2787K}
Koz{\l}owski, S. 2016, MNRAS, 459, 2787

\bibitem[Lomb(1976)]{1976Ap&SS..39..447L} 
Lomb, N. R. 1976, Ap\&SS, 39, 447

\bibitem[MAGIC Collaboration et al.(2023)]{2023A&A...670A..49M} 
MAGIC Collaboration, Acciari, V. A., Aniello, T., et al. 2023, A\&A, 670, A49

\bibitem[Mannheim et al.(2020)]{2020NatCo..11.4176M}
Mannheim, K., et al. 2020, Nature Communications, 11, 4176

\bibitem[Marscher(2014)]{2014ApJ...780...87M} 
Marscher, A. P. 2014, ApJ, 780, 87

\bibitem[Marscher \& Gear(1985)]{1985ApJ...298..114M} 
Marscher, A. P., \& Gear, W. K. 1985, ApJ, 298, 114

\bibitem[Masci et al.(2019)]{2019EPSC...13.1163M} 
Masci, F., Ye, Q., Kramer, E. A., et al. 2019, EPSC, 13, 1163

\bibitem[Massaro et al.(2001)]{2001A&A...374..435M} 
Massaro, E., Mantovani, F., Fanti, R., et al. 2001, A\&A, 374, 435

\bibitem[Negi et al.(2023)]{2023MNRAS.524L..66N} 
Negi, V., Gopal-Krishna, Chand, H., \& Britzen, S. 2023, MNRAS, 524, L66

\bibitem[Negi et al.(2022)]{2022MNRAS.516.5376N}
Negi, V., Joshi, R., Gaur, H., et al. 2022, MNRAS, 516, 5376

\bibitem[Nilsson et al.(2018)]{2018A&A.620A.185N} 
Nilsson, K., Lindfors, E., Takalo, L. O., et al. 2018, A\&A, 620, A185

\bibitem[Paliya et al.(2017)]{2017ApJ...844...32P}
Paliya, V.~S., Stalin, C.~S., Ravikumar, C.~D., et al.\ 2017, ApJ, 844, 32

\bibitem[Pandey et al.(2022)]{2022MNRAS.510.1809P} 
Pandey, A., Rajput, B., \& Stalin, C. S. 2022, MNRAS, 510, 1809

\bibitem[Pandian et al.(2022)]{2022JApA...43...48P}
Pandian, K.~S.~U., Natarajan, A., Stalin, C.~S., Muneer, S., Natarajan, B. 2022, J.\ Astrophys.\ Astr., 43, 48

\bibitem[Perlman et al.(2005)]{2005ApJ...625..727P} 
Perlman, E. S., Madejski, G., Georganopoulos, M., et al. 2005, ApJ, 625, 727

\bibitem[Peterson et al.(2004)]{2004ApJ...613..682P} 
Peterson, B. M., Ferrarese, L., Gilbert, K. M., et al. 2004, ApJ, 613, 682

\bibitem[Peterson et al.(1998)]{1998PASP..110..660P} 
Peterson, B. M., Wanders, I., Horne, K., et al. 1998, PASP, 110, 660

\bibitem[Press(1978)]{1978ComAp...7..103P} 
Press, W. H. 1978, ComAp, 7, 103

\bibitem[Piner \& Edwards(2014)]{2014ApJ...797...25P} 
Piner, B. G., \& Edwards, P. G. 2014, ApJ, 797, 25

\bibitem[Raiteri et al.(2017b)]{2017Natur.552..374R} 
Raiteri, C. M., Villata, M., Acosta-Pulido, J. A., et al. 2017b, Nature, 552, 374

\bibitem[Raiteri et al.(2017a)]{2017SSRv..207....5R} 
Raiteri, C. M., Villata, M., Cheng, H. W., et al. 2017a, Space Sci. Rev., 207, 5

\bibitem[Raiteri et al.(2003)]{2003A&A...402..151R} 
Raiteri, C. M., Villata, M., Tosti, G., et al. 2003, A\&A, 402, 151

\bibitem[Rajput et al.(2022)]{2022MNRAS.517.3236R} 
Rajput, B., Pandey, A., Stalin, C. S., \& Mathew, B. 2022, MNRAS, 517, 3236

\bibitem[Rakshit et al.(2017)]{2017ApJ...835..275R} 
Rakshit, S., Stalin, C. S., Muneer, S., et al. 2017, ApJ, 835, 275

\bibitem[Scargle(1982)]{1982ApJ...263..835S} 
Scargle, J. D. 1982, ApJ, 263, 835

\bibitem[Schlafly \& Finkbeiner(2011)]{2011ApJ...737..103S} 
Schlafly, E. F., \& Finkbeiner, D. P. 2011, ApJ, 737, 103

\bibitem[Sesar et al.(2007)]{2007AJ....134.2236S} 
Sesar, B., Ivezi{\'c}, {\v{Z}}., Lupton, R. H., et al. 2007, AJ, 134, 2236

\bibitem[Simonetti et al.(1985)]{1985ApJ...296...46S} 
Simonetti, J. H., Cordes, J. M., \& Heeschen, D. S. 1985, ApJ, 296, 46

\bibitem[Tody(1986)]{1986SPIE..627..733T} 
Tody, D. 1986, SPIE, 627, 733

\bibitem[Urry \& Padovani(1995)]{1995PASP..107..803U} 
Urry, C. M., \& Padovani, P. 1995, PASP, 107, 803

\bibitem[Uttley et al.(2002)]{2002MNRAS.332..231U} 
Uttley, P., McHardy, I. M., Papadakis, I. E. 2002, MNRAS, 332, 231

\bibitem[Vagnetti et al.(2003)]{2003ApJ...590..123V} 
Vagnetti, F., Trevese, D., \& Nesci, R. 2003, ApJ, 590, 123

\bibitem[Vagnetti et al.(2011)]{2011A&A...536A..84V} 
Vagnetti, F., Turriziani, S., Trevese, D. 2011, A\&A, 536, A84

\bibitem[Vanden Berk et al.(2004)]{2004ApJ...601..692V}
Vanden Berk, D.~E., Wilhite, B.~C., Kron, R.~G., et al. 2004, ApJ, 601, 692

\bibitem[Vaughan et al.(2003)]{2003MNRAS.339.1237V} 
Vaughan, S., Fabian, A. C., \& Nandra, K. 2003, MNRAS, 339, 1237

\bibitem[Wagner \& Witzel(1995)]{1995ARA&A..33..163W} 
Wagner, S. J., \& Witzel, A. 1995, ARA\&A, 33, 163

\bibitem[Wandel et al.(1999)]{1999ApJ...526..579W} 
Wandel, A., Peterson, B. M., \& Malkan, M. A. 1999, ApJ, 526, 579

\bibitem[Welsh et al.(2011)]{2011PASP..123..634W}
Welsh, W.~F., Wheatley, J.~M., Neilson, H.~R., et al. 2011, PASP, 123, 634

\bibitem[Wiita(1996)]{1996ASPC..110...42W} 
Wiita, P. J. 1996, ASPC, 110, 42
\end{thebibliography}
\end{document}